\documentclass[aps,amsfonts,amsmath,amssymb,pra,showpacs]{revtex4}

\usepackage{graphicx}
\newcommand{\be}{\begin{equation}}
\newcommand{\ee}{\end{equation}}
\newcommand{\bex}{\begin{eqnarray}}
\newcommand{\eex}{\end{eqnarray}}
\newcommand{\bsub}{\begin{subequations}}
\newcommand{\ensub}{\end{subequations}}
\newtheorem{thm}{Theorem}
\newtheorem{crl}{Corollary}
\def\qed{$\Box$}
\newtheorem{defni}{Definition}

\begin{document}
\title{Generalized Quantum Secret 
Sharing} % \footnote{\large Accepted for publication in Physical Review A}}
\author{Sudhir Kumar Singh}
\email{suds@ee.ucla.edu}
\affiliation{Department of Electrical Engineering, University of California, Los Angeles,
CA 90095}
\author{R. Srikanth}
\email{srik@rri.res.in}
\affiliation{Raman Research Institute, Bangalore- 560080, India}
\begin{abstract}
We explore a generalization of quantum secret sharing (QSS) in which classical shares
play a complementary role to quantum shares, exploring further consequences
of an idea first studied by Nascimento, Mueller-Quade and Imai
(Phys. Rev. {\bf A64} 042311 (2001)). We examine three ways, termed inflation,
compression and twin-thresholding, by which the proportion of classical shares can be
augmented. This has the important application that it reduces quantum (information
processing) players by replacing them with their classical counterparts, 
thereby making quantum secret sharing
considerably easier and less expensive to implement in a practical setting.
In compression, a QSS scheme is turned into an equivalent scheme with fewer quantum
players, compensated for by suitable classical shares. 
In inflation, a QSS scheme is enlarged by adding only classical shares and players. 
In a twin-threshold scheme, we invoke two separate thresholds for classical and quantum
shares based on the idea of information dilution.
\end{abstract}

\maketitle

\section{Introduction:}\label{intro}

Suppose the president of a bank, Alice, wants to give access to a vault to two 
vice-presidents, Bob and Charlie, whom she does not entirely trust.
Instead of giving the combination to any one of them, she may desire to
distribute the information in such a way that no vice-president alone has any 
knowledge of the combination, but both of them can jointly determine the combination.
Cryptography provides the answer to this question in the form of
{\it secret sharing} \cite{schneier96,gruska97}.
In this scheme, some sensitive data is distributed among a
number of parties such that certain authorized sets of parties can 
access the data, but no other combination of players.
A particularly symmetric variety of secret splitting (sharing) is
called a {\it threshold scheme}: 
in a $(k,n)$ classical threshold scheme (CTS),
the secret is split up into $n$ pieces (shares),
of which any $k$ shares form a set {\em authorized} to reconstruct the secret, while any set of 
$k-1$ or fewer shares has no information about the secret.
Blakely \cite{blakely79} and Shamir \cite{sha79} showed that CTS's exist
for all values of $k$ and $n$ with $n \geq k$. By concatenating threshold schemes,
one can construct arbitrary access structures, subject only to the condition of
monotonicity (ie., sets containing authorized sets should also be authorized)
\cite{ben90}. Hillery {\em et al.}
\cite{hil00} and Karlsson {\em et al.} \cite{kar00} proposed methods for
implementing CTSs that use {\em quantum} information
to transmit shares securely in the presence of eavesdroppers.

Subsequently, extending the above idea to the quantum case,
Cleve, Gottesman and Lo \cite{cle00} proposed
 a $(k,n)$ {\it quantum} threshold scheme (QTS) as a method to split up 
an unknown secret quantum state $|S\rangle$ into $n$ 
pieces (shares) with the restriction that 
% Each share can consist of any number of qubits (or higher-dimensional states),
% and not all shares need to be of the same size.
$k > n/2$ (for if this inequality were violated, two disjoint sets of players
can reconstruct the secret, in violation of the quantum no-cloning theorem \cite{woo82}). 
The notion of QTS is based on quantum erasure correction \cite{cs,gra97}.
QSS has been extended beyond QTS to
general access structures \cite{got00,smi00}, but here the no-cloning theorem implies
that none of the authorized sets shall be mutually disjoint.
Potential applications of QSS include
creating joint checking accounts containing quantum money \cite{wiesner83}, 
or share hard-to-create ancilla
states \cite{got00}, or perform secure distributed quantum computation \cite{cre01}.
Implementing QSS is well within current technology \cite{lan03}, and has been
demonstrated by a recent experiment \cite{lan04}.

In conventional QSS schemes, it is often implicitly assumed that 
all share-holders carry and process quantum information. Given the fragile nature
of quantum information, this can often be difficult and expensive from a practical
viewpoint. 
Fortunately, it turns out to be possible sometimes to construct an equivalent scheme in 
which some share holders carry only classical information and no
quantum information, an idea first studied by Nascimeto et al. \cite{nas01}. 
Our work is dedicated to exploring further consequences of this idea.
It is of considerable importance to consider 
ways in which the proportion of classical shares and classical information processing can
be increased in realizing a QSS scheme. 
Furthermore, hybrid (classical-quantum) QSS can potentially avail of features
available to classical secret sharing such as share renewal \cite{herzberg},
secret sharing with prevention \cite{beltel} and disenrolment \cite{martin}.

In particular, our work is aimed at studying ways to augment the proportion of classical 
shares in different ways for various situations in QSS.  As pointed out above, the main purpose
of this exercise is to render practical implementation easier and less expensive.
In Section \ref{sec:hybri}, we present some
basic ways to introduce a classical information component into QSS.
In Section \ref{sec:compress}, we discuss how this can be used to `compress' a QSS, that is,
reduce the proportion of quantum information carrying players.
In Section \ref{sec:infla}, we show how a QSS scheme can be `inflated'
by adding only classical shares. 
In Section \ref{sec:qts}, we invoke two separate thresholds for classical and quantum
shares based on the idea of information dilution. This generalizes the idea of conventional
single threshold QSS schemes and is again shown to lead to savings of quantum players.

\section{Hybridizing quantum secret sharing schemes}\label{sec:hybri}

The essential method to hybridize (i.e., to introduce classical shares into) QSS is to somehow
incorporate classical information that
is needed to decrypt or prepare the quantum secret as classical shares. 
Use of classical shares can sometimes obviate and thus lead to savings in
quantum shares, or, at any rate, quantum players.
A simple instance of such classical information is the ordering information 
of the shares. In QTS, it is implicity assumed that the share-holders know 
the coordinates of the shares in the secret, i.e., they know who is holding the
first qubit, who the second and so on. This ordering
information is necessary to reconstruct the secret, without which successful
reconstruction of the secret is not guaranteed. If we wish to make use of this
ordering information in the above sense, then only quantum error correction 
based secret sharing where lack of ordering information leads to maximal
ignorance can be used. In particular, the scheme should be 
sensitive to the interchange of two or more qubits.
For example, let us consider a $(2,3)$-QTS. The secret here is an arbitrary qutrit
and the encoding maps the secret qutrit to three qutrits as:
\begin{equation}
\alpha|0\rangle + \beta|1\rangle + \gamma|2\rangle \longmapsto
\alpha(|000\rangle + |111\rangle + |222\rangle) + 
\beta(|012\rangle + |120\rangle + |201\rangle) + 
\gamma(|021\rangle + |210\rangle + |102\rangle),
\end{equation}
and each qutrit is taken as a share. While from a single share no information can
be obtained, two shares, with ordering information, suffice to reconstruct the
encoded state \cite{cle00}. However, the lack of ordering information does not
always lead to maximal ignorance about the secret. Note that the structure of the above
code is such that any interchange of two qubits  leaves an encoded $|0\rangle$
intact but interchanges $|1\rangle$ and $|2\rangle$. Thus, a secret like $|0\rangle$
or $(1/\sqrt{2})(|1\rangle + |2\rangle)$ can be entirely
reconstructed without the ordering
information. Therefore, only the subset of quantum error correction codes admissible
in QSS that do not possess such symmetry properties can be used if the scheme is
to be sensitive to ordering information.

Another scheme relevant here is due to Nascimento et al. \cite{nas01},
based on qubit encryption \cite{qcrypt}. 
% In Sections \ref{sec:infla} and \ref{sec:compress_aqss},
We adopt this method to generate the relevant encrypting classical information.
However, in principle any classical data
whose suppression leads to maximal ignorance of the secret
is also good. Elsewhere, in Section \ref{sec:qts}, we consider another way.
Quantum encryption works as follows: suppose we have a 
$n$-qubit quantum state $|\psi\rangle$ and random sequence $K$ of $2n$
classical bits. Each sequential pair of classical bit is associated with a qubit
and determines which transformation $\hat{\sigma} \in \{\hat{I}, \hat{\sigma}_x,
\hat{\sigma}_y, \hat{\sigma}_z\}$ is applied to the respective qubit. If the
pair is 00, $\hat{I}$ is applied, if it is $01$, $\hat{\sigma}_x$ is applied,
and so on. To one not knowing $K$, 
the resulting $|\tilde{\psi}\rangle$ is a complete mixture and
no information can be extracted out of it because the encryption leaves
any pure state in a maximally mixed state, that is: 
$(1/4)(\hat{I}|S\rangle\langle S|\hat{I} +
\hat{\sigma}_x|S\rangle\langle S|\hat{\sigma}_x +
\hat{\sigma}_y|S\rangle\langle S|\hat{\sigma}_y +
\hat{\sigma}_z|S\rangle\langle S|\hat{\sigma}_z) = (1/2)\hat{I}$.
However, with knowledge of $K$
the sequence of operations can be reversed 
and $|\psi\rangle$ recovered. Therefore, classical data can be used
to encrypt  quantum data. In general, given $d$ dimensional objects,
quantum encryption requires $d^2$ operators and a key of $2\log(d)$ bits per object
to randomize perfectly. In practice, such quantum operations may prove costly,
and only near-perfect security may be sufficient. In this case,
there exists a set of roughly $d\log(d)$ unitary operators whose average effect on every 
input pure state is almost perfectly randomizing, 
so that the size of the key can be reduced by about a factor of two \cite{hay03}.

\section{Compressing Quantum Secret Sharing Schemes}\label{sec:compress}

In hybrid QSS, the quantum secret is split up into quantum and classical shares of
information. We call the former q-shares, and the latter c-shares. A player holding only
c-shares is called a c-player or c-member. Otherwise, she or he is a q-player
or q-member. 
\begin{defni}
A QSS scheme realizing an access structure $\Gamma = \{\alpha_1, \alpha_2,
\cdots,\alpha_r\}$ among a set of players ${\cal P} = \{P_1, P_2, \cdots, P_n\}$
is said to be compressible if fewer than $n$ q-players are sufficient to implement it. 
\end{defni}
Here the $\alpha_i$'s are the minimal authorized sets of players.
Knowledge of compressibility
helps us decide how to minimize valuable quantum  resources needed for 
implementing a given QSS scheme. 
As an example of compression by means of hybrid QSS, suppose we want to split a
quantum secret $|S\rangle$ among a set of players ${\cal P} = \{A, B, C, D, E, F\}$
realizing the access structure $\Gamma = \{ ABC, AD, AEF\}$. That is, the only sets 
authorized to reconstruct the secret are $\{A,B,C\}$, $\{A,D\}$ and $\{A,E,F\}$
and sets containing them, whilst any other set is unauthorized to do so. For distributing
the secret, we encrypt
$|S\rangle$ using the quantum encryption method (described above) with classical key
$K$ into a new state $|\tilde{S}\rangle$ and give $|\tilde{S}\rangle$ to $A$.
We then split up $K$ using a CSS scheme that realizes $\Gamma$. Player $A$ cannot
recover $|S\rangle$ from $|\tilde{S}\rangle$ because he cannot unscramble it without
$K$. Only the $\alpha_j$'s, and sets containing them, can recover the
classical key $K$, and thence decrypt the secret state. In this way, by means
of a hybrid (classical-quantum) secret-sharing scheme, we can compress the original
QSS scheme into an equivalent one in which fewer players need to handle quantum
information. 

We use the notation of single parantheses (double parantheses) to indicate
CTS (QTS). In a conventional $((k,n))$ scheme, a compression is known to be possible
if $2k > n+1$, in which case, the scheme can be compressed into a
$((k-\gamma,n-\gamma))$ scheme combined with a $(k,n)$ scheme, where
$\gamma \equiv 2k - n - 1$ \cite{nas01}. A general access structure
$\Gamma = \{\alpha_1, \alpha_2,\cdots, \alpha_r\}$ can be realized by a first layer of
$(1,r)$-threshold scheme. In the quantum case, since this violates the no-cloning
theorem, it is replaced by the majority function $((r,2r-1))$ scheme \cite{got00}.
This, again, is incompressible. However, in the second layer of the construction, 
the $((|\alpha_i|,|\alpha_i|))$ schemes can be replaced with a $((1,1))$ schemes combined
with $(|\alpha_i|,|\alpha_i|)$ schemes \cite{nas01}. 

In the above, the degree of compression is determined by $\Gamma = 
\{\alpha_1,\cdots,\alpha_r\}$ and the requirement to
minimize q-players, {\em no matter who they are}. This can be distorted if the information
processing capabilities of individual players are known to be different. In particular,
suppose we are given a set $\mathbb{Q}$, such that players from this set are able
to process quantum information reliably. 
The set of remaining players $\bar{\mathbb{Q}} = {\cal P} - \mathbb{Q}$
are best designated to be c-players.
A `hitting set' ${\cal H}(\Gamma)$ for the collection of
sets $\Gamma$ is a set of players such that ${\cal H} \cap \alpha_i \ne \emptyset 
~\forall~ i ~(1 \le i \le r)$. Let ${\cal M}(\Gamma)$ be the smallest
hitting set for $\Gamma$ such that ${\cal M}(\Gamma) \subset \mathbb{Q}$. 
${\cal M}(\Gamma)$ may or may not be unique. We denote $M \equiv |{\cal M}(\Gamma)|$. 
Under compression, only $M$ q-players are needed. First a majority function
$((r,2r-1))$ is employed, $r$ of the shares being encrypted and deposited with the $M$
q-players. In the second layer of the construction,
the $((|\alpha_i|,|\alpha_i|))$ schemes can be compressed to $((1,1))$ schemes combined
with $(|\alpha_i|,|\alpha_i|)$ schemes, the q-shares of each $\alpha_i$ being deposited with 
the respective q-player.
The remaining $M-1$ shares are split-shared according to a maximal scheme that contains
$\Gamma$. The maximal scheme is obtained by adding authorized sets to 
$\Gamma$ until authorized and unauthorized sets form exact complements \cite{got00}. 
Thus we require only $M \le |{\cal P}|$ q-players to implement the protocol,
which represents a compression by $|{\cal P}|-M$ q-players. We note that if
${\cal M}(\Gamma) = \emptyset$, then compression is impossible.
We can observe that 
for a general access structure involving a large number of players, computing
${\cal M}$ is a provably hard problem (In fact its decision version can be shown to
be NP-Complete \cite{karp72, garey79}).

As an example, let us consider the access structure
$\Gamma = \{ABCD, ADF, CDE\}$ among six players ${\cal P} = \{A, B, C, D, E, F\}$. 
Suppose $\mathbb{Q} = \{A, C, E\}$.
We can choose ${\cal M} = \{A, C\}$ or ${\cal M} = \{A, E\}$, 
representing the two required q-players (instead of six q-players,
required in the uncompressed version). Suppose we choose the latter.
The first layer will employ a $(3,5)$-QTS to split secret $|S\rangle$.
At the second layer, the share on the top two rows
are encrypted using $K_1, K_2$ and given 
to $A$, the last using $K_3$ and given to $E$. The $K_j$'s are classically shared 
on each row, though the q-shares remains with $A$ or $E$. The last share $|S^{\prime}\rangle$
is shared using a pure state scheme that implements $\Gamma_{\max}$, the maximal
scheme obtained from $\Gamma$.
% unauthorized sets form exact complements) \cite{got00}.
The resultant q-shares for each authorized set $\alpha_j$ are deposited with
$\alpha_j \cap {\cal M}(\Gamma)$. This scheme is depicted in Eq. (\ref{eq:qz1}). 
\begin{equation}
\label{eq:qz1}
((3,5)) \left\{ \begin{array}{lll}
A & \rightarrow (4,4) : & A, B, C, D \\
A & \rightarrow (3,3) : & A, D, F\\
E & \rightarrow (3,3) : & C, D, E \\
|S^{\prime}\rangle &  &  
	   \end{array} \right. 
\end{equation}
We note that if $D \in \mathbb{Q}$ then ${\cal M}(\Gamma) = \{D\}$,
and only one q-player, namely $D$, would have sufficed to implement compression. 
And if $\mathbb{Q} = \{E, F\}$, then ${\cal M}=\emptyset$ because there is an
authorized set with no q-player. Hence no compression would be possible.

\section{Inflating Quantum Secret Sharing Schemes}\label{sec:infla}

The question, how to augment or ``inflate"
a given QSS scheme keeping
the quantum component fixed, is considered herebelow. This is of practical relevance
if we wish to expand a given QSS scheme
by including new players who do not
have (reliable) quantum information processing capacity. To this end, we now define
an inflatable QSS.
\begin{defni}
A QSS($\Gamma$) scheme
realizing an access structure $\Gamma = \{\alpha_1, \alpha_2,
\cdots,\alpha_r\}$ among a set of players ${\cal P} = \{P_1, P_2, \cdots, P_n\}$
% using a total of $m$ q-shares 
is inflatable if $n$ can be increased for fixed number of
q-players to form a new QSS($\Gamma^{\prime}$)
such that $\Gamma^{\prime}|_{\cal P} = \Gamma$,
where $\Gamma^{\prime}|_{\cal P}$ denotes the 
restriction of $\Gamma^{\prime}$ to ${\cal P}$.
\end{defni}
Clearly, inflation involves the addition of classical
information carrying c-players. The additional
shares required for them will be c-shares,
so that q-shares may remain fixed at $m$.
The following theorem answers the question when a QSS scheme
can be inflated.

\begin{thm}
\label{thm:alinf}
A QSS scheme realizing an access structure $\Gamma = \{\alpha_1, \alpha_2,
\cdots,\alpha_r\}$ among a set of players ${\cal P} = \{P_1, P_2, \cdots, P_n\}$
% using a total of $m$ q-shares 
can always be inflated.
\end{thm}

\noindent
{\bf Proof.} Consider the addition of $m$ new players, $P_{n+1},\cdots P_{n+m}$,
where $m \ge 1$.
The new set of all players are ${\cal P}^{\prime} \equiv
\{P_1, P_2, \cdots,P_{n+m}\}$. A new access
structure $\Gamma^{\prime}=\{\alpha_1^{\prime},\cdots,\alpha_r^{\prime}\}$ 
can be obtained by arbitrarily adding the new players
to any of the $\alpha_j$'s. Clearly, $\Gamma^{\prime}$ will not violate the
no-cloning theorem, since $\Gamma$ does not. The secret $|S\rangle$ is encrypted using
classical string $K$ to obtain $|\tilde{S}\rangle$. This encrypted secret is split-shared
according to the original scheme implementing $\Gamma$, while $K$ is split-shared
among all $n+m$ players according to the classical scheme implementing $\Gamma^{\prime}$.
To reconstruct the secret, members of any $\alpha_j^{\prime}$ combine the
q-shares of $\alpha_j \subseteq \alpha_j^{\prime}$ 
to reconstruct $|\tilde{S}\rangle$, and the c-shares with all
members of  $\alpha_j^{\prime}$ to reconstruct $K$, using which the encrypted secret 
$|\tilde{S}\rangle$ is decrypted to $|S\rangle$.
The new scheme is such that $\Gamma = \Gamma^{\prime}|_{\cal P}$ and
${\cal P}^{\prime}-{\cal P}$ are c-players.
% the no-cloning theorem,
% where $\Gamma^{\prime}|_{{\cal P}^{\prime}-P_i}$ denotes the 
% restriction of $\Gamma^{\prime}$ to ${\cal P}^{\prime} - P_i$.
% As a result, the augmented 
% scheme can be realized as a conventional QSS scheme using (say) $m^{\prime} > m$ q-shares.
% Therefore, by construction, there is one $P_i$
% (namely, that for $i=n+1$ in the above case) 
% such that $\Gamma^{\prime}|_{{\cal P}^{\prime}-P_i}$ does not violate
% the no-cloning theorem,
% where $\Gamma^{\prime}|_{{\cal P}^{\prime}-P_i}$ denotes the 
% restriction of $\Gamma^{\prime}$ to ${\cal P}^{\prime} - P_i$.
% Therefore, according to Theorem 1 of Ref. \cite{nas01}, the new QSS scheme obtained
% by adding $P_{n+1}$ is compressible, meaning that $P_{n+1}$ can be a c-player.
Therefore, the new scheme QSS($\Gamma^{\prime}$) is an inflation of 
the given scheme QSS($\Gamma$).  \hfill \qed
\bigskip

The above theorem only says that that any QSS scheme can be inflated in {\em some}
way. A specific problem is whether a given $(k,n)$-QTS  can be inflated to another QTS.
This is considered in the following theorem and corollary.

% \noindent {\bf Proof.} Suppose $(k,n)$-QTS can be inflated at constant threshold. Then
% there exists a $(k,n^{\prime})$-QTS, consistent with the no-cloning theorem
% and with $n^{\prime}>n$, whose restriction leads to $(k,n)$-QTS.
% Let $n^{\prime} - n \equiv \gamma$. The restriction of $(k,n^{\prime})$-QTS
% by $\gamma$ players will lead to a $(k-\gamma,n)$-QTS \cite{nas01}, 
% whose access structure
% is different from $(k,n)$-QTS. This contradicts our original assumption. 
% \hfill \qed
% \bigskip

\begin{thm}
\label{thm:infla}
A $(k,n)$-QTS can be inflated only conformally, ie., to 
threshold schemes having the form $(k+\gamma,n+\gamma)$,
where $\gamma$ $(\ge 0)$ is an integer. \end{thm}

\noindent
{\bf Proof.} If the given $(k,n)$-QTS satisfies the no-cloning theorem, then
clearly so will the $(k+\gamma_k,n+\gamma_n)$-QTS, where $\gamma_k \ge \gamma_n \ge 0$ and
$k+\gamma_k \le n+\gamma_n$. Further, according to Lemma 1
of Ref. \cite{nas01}, a restriction of the $(k+\gamma_k,n+\gamma_n)$-QTS by $\gamma$ players 
necessarily yields a conformally reduced, $(k+\gamma_k-\gamma,n+\gamma_n-\gamma)$-QTS. 
The restricted scheme has a different access structure from $(k,n)$-QTS unless
$\gamma_k=\gamma_n=\gamma$. Therefore, only a conformal inflation of $(k,n)$-QTS is
possible, whereby it is inflated to a $(k+\gamma,n+\gamma)$-QTS 
by the addition of $\gamma$ c-players.
\hfill \qed
\bigskip

In an implementation of Theorem \ref{thm:infla}, the quantum secret 
$|S\rangle$ is encrypted to $|\tilde{S}\rangle$ using a
classical string $K$ which is split-shared among all $n+\gamma$ players
according to a \mbox{$(k+\gamma,n+\gamma)$-CTS}. State $|\tilde{S}\rangle$ is then 
quantally split-shared among the $n$ q-players according to a \mbox{$(k,n)$-QTS}.
As a consequence of Theorem \ref{thm:infla}, we have the following negative result:
\begin{crl}
A $(k,n)$-QTS cannot be inflated at constant threshold.
\hfill \qed
\end{crl}

\section{Twin-threshold quantum secret sharing schemes}\label{sec:qts}
In a conventional or compressed $(k,n)$-QTS, the threshold $k$ applies to
all members taken together. Now suppose that we have {\em separate} thresholds
for c-members and q-members, namely $k_c$ and $k_q$, with $k=k_c+k_q$. 
We now extend the definition of a conventional QTS 
to a $(k_c,k_q,n)$ {\it quantum twin-threshold scheme} (Q2TS) 
and a $(k_c,k_q,n,\mathbb{C})$
{\it quantum twin-threshold scheme with common set} (Q2TS+C), 
where a quantum secret $|S\rangle$
is split into $n$ pieces (shares) according to some pre-agreed procedure
and distributed among $n$ players. 
These $n$ share-holders consist of members
of set $\mathbb{Q}$ of q-players
and set $\bar{\mathbb{Q}}$ of c-players. 
We denote $q \equiv |\mathbb{Q}|$, so that $|\bar{\mathbb{Q}}| = n-q$.
Obviously, in a quantum scheme, $\mathbb{Q} \ne \emptyset$.

\begin{defni}
A QSS scheme is a $(k_c,k_q,n)$ quantum twin-threshold scheme (Q2TS) 
among $n$ players, of which $q$ are q-players and the remaining are c-players,
if at least $k_c$ c-players and at least $k_q$ q-players are necessary to
reconstruct the secret.
\end{defni}

\begin{defni}
A QSS scheme is a  $(k_c,k_q,n,\mathbb{C})$ quantum 
twin-threshold scheme with common set (Q2TS+C)
among $n$ players, of which $q$ are q-players and the remaining are c-players,
if: (a) at least $k_c$ c-players and at least $k_q$ q-players are necessary to
reconstruct the secret; 
(b) All members of the set $\mathbb{C}$ are necessary to reconstruct the secret. 
\end{defni}

The idea behind distinguishing between the classical threshold $k_c$
and the quantum threshold $k_q$ is to obtain a 
simple generalization that combines the properties of the
CTS and QTS. 
Practically speaking, it is best to minimize $k_q$, at fixed $k$. 
However, one can in principle consider situations of potential use for a
twin-threshold scheme, when a sufficiently large number of members are 
able to process quantum information safely.
Further, some of the share-holders, while not entirely trust-worthy, may yet
be more trust-worthy than others. The share-dealer (say Alice) may prefer
to include all such share-holders during any reconstruction of the secret.
This is the requirement that motivates the introduction of set $\mathbb{C}$.
In general, $\mathbb{C}$ can contain members drawn from $\mathbb{Q}$
and/or $\bar{\mathbb{Q}}$ or may be a null set. 
By definition, Q2TS+C with $\mathbb{C} = \emptyset$ is Q2TS.

In the following sections we present two methods to realize
in varying degrees the generalized quantum secret splitting scheme.
The first of these is the general version of Q2TS+C. The second, while
more restricted, is interesting
because it is not directly based on quantum erasure correction, but on
information dilution via homogenization, in contrast to current proposals of QSS. 

\subsection{Quantum error correction and quantum encryption}

We now give protocols that realizes the twin-threshold scheme based on 
quantum encryption.

{\em Scheme 1.} Protocol to realize $(k_c,k_q,n)$-Q2TS.
 
{\em Distribution phase.} (1) Choose a random classical encryption $K$. Encrypt the
quantum secret $|S\rangle$ using the encryption algorithm described in Section
\ref{intro}. The encrypted state is denoted $|\tilde{S}\rangle$; (2) Using a conventional
$(k_q, q)$-QTS, split-share $|\tilde{S}\rangle$ among the members of $\mathbb{Q}$;
to not violate no-cloning, $q$ should satisfy $k_q > (q/2)$;
(3) Using a $(k_c,n-q)$-CTS, split-share $K$ among the members of $\bar{\mathbb{Q}}$.

{\em Reconstruction phase.} (1) Collect any $k_q$ q-shares from members of
$\mathbb{Q}$ and reconstruct 
$|\tilde{S}\rangle$; (2) Collect any $k_c$ shares from members of $\bar{\mathbb{Q}}$
and reconstruct $K$;
(3) Reconstruct $|S\rangle$ using $|\tilde{S}\rangle$ and $K$.

Now consider the case $\mathbb{C}
\ne \emptyset$ and the Q2TS scheme becomes the more general Q2TS+C scheme.
We now give a protocol that realizes this more general twin-threshold scheme.
We denote $\lambda_q \equiv |\mathbb{Q} \cap \mathbb{C}|$ and 
$\lambda_c \equiv |\bar{\mathbb{Q}} \cap \mathbb{C}|$. Clearly, $\lambda_c +
\lambda_q = |\mathbb{C}|$. 
If there are no q-players in $\mathbb{C}$, set $\lambda_q
= 0$, and if there are no c-players in $\mathbb{C}$, set $\lambda_c = 0$.
Note that by definition, q-players may also carry classical information, but
c-players don't carry quantum information.

{\em Scheme 2.} Protocol to realize $(k_c,k_q,n,\mathbb{C})$-Q2TS+C.
 
{\em Distribution phase.} (1) Choose a random classical encryption $K$. Encrypt the
quantum secret $|S\rangle$ using the encryption algorithm described in Section
\ref{intro}. The encrypted state is denoted $|\tilde{S}\rangle$; 
(2) Using a $(2,2)$-QTS, divide $|\tilde{S}\rangle$ into two pieces, say
$|\tilde{S}_1\rangle$ and $|\tilde{S}_2\rangle$; 
(3) Using a $(\lambda_q,\lambda_q)$-QTS, split $|\tilde{S}_1\rangle$ 
among the q-members in $\mathbb{C}$;
(4) Using a conventional
$(k_q-\lambda_q, q-\lambda_q)$-QTS, split $|\tilde{S}_2\rangle$ among the 
q-members not in $\mathbb{C}$;
to not violate no-cloning, $q$ should satisfy $(k_q-\lambda_q) > (q-\lambda_q)/2$;
(5) Using a (2,2)-CTS, divide $K$ into two shares, say $K_1$ and $K_2$;
(6) Part $K_1$ is split among the members of $\mathbb{C}$ using a
$(|\mathbb{C}|,|\mathbb{C}|)$-CTS. Alternatively, it can be split using a
$(\lambda_c,\lambda_c)$-CTS among the c-players in $\mathbb{C}$;
(7) Using a $(k_c-\lambda_c,n-q-\lambda_c)$-CTS, 
split $K_2$ among the members of $\bar{\mathbb{Q}}-\mathbb{C}$.

{\em Reconstruction phase.} 
(1) Collect all $\lambda_q$ shares from all members of $\mathbb{Q} \cap \mathbb{C}$
and reconstruct $|\tilde{S}_1\rangle$; 
(2) Collect any $k_q-\lambda_q$ q-shares from
$\mathbb{Q} -\mathbb{C}$ to reconstruct $|\tilde{S}_2\rangle$;
(3) Combining $|\tilde{S}_1\rangle$ and $|\tilde{S}_2\rangle$, reconstruct
$|\tilde{S}\rangle$;
(4) Collect all $|\mathbb{C}|$ c-shares 
from members of $\mathbb{C}$ and reconstruct $K_1$.
Alternatively, collect all $\lambda_c$ c-shares 
from members of $\bar{\mathbb{Q}} \cap \mathbb{C}$ and reconstruct $K_1$; 
(5) Collect any $k_c-\lambda_c$ shares from $\bar{\mathbb{Q}}-\mathbb{C}$
and reconstruct $K_2$; 
(6) Combining $K_1$ and $K_2$, reconstruct $K$.
(7) Reconstruct $|S\rangle$ using $|\tilde{S}\rangle$ and $K$.

\subsection{Quantum twin-threshold scheme based on information dilution via homogenization}
The second, more restrictive scheme, is based on the procedure for information 
dilution in a system-reservoir 
interaction, proposed by Ziman {\it et al.} \cite{zim02}. The novelty of the
scheme lies in the fact that it is not directly based on an quantum error-correction
code. However, it is applicable only to QSS with $\mathbb{C} \ne \emptyset$.
Ref. \cite{zim02} present a {\it universal quantum homogenizer}, a machine
that takes as input a system qubit initially in the state 
$\rho$ and a set of $N$ reservoir qubits initially 
prepared in the identical state $\xi$.
In the homogenizer the system qubit sequentially interacts with the
reservoir qubits via the partial swap operation so that the initial state
$\rho^{(0)}_S$ of the system, after interacting with the $N$ reservoir qubits, becomes:
\begin{equation}
\rho^{(N)} = {\rm Tr}_R\left[U_N\cdots U_1(\rho_S^{(0)}\otimes\xi^{\otimes N})
U_1^{\dag}\cdots U^{\dag}_N\right]
\end{equation}
where $U_k \equiv U\otimes (\otimes_{j\ne k}\mathbb{I}_j)$ describes the interaction between
the $k$th qubit of the reservoir and the system qubit.
The homogenizer realizes, in the limit sense, 
the transformation such that at the
output each qubit is in an arbitrarily small
neighborhood of the state $\xi$ irrespective of
the initial states of the system and the reservoir qubits. Formally,
\bsub
\bex
\label{secu}
D(\rho_S^{(N)},\xi) \le \delta,~~~~\forall N\ge N_{\delta}, \label{secu1} \\
D(\xi^{\prime}_k,\xi) \le \delta,~~~~\forall 1 \le k \le N, \label{secu2}
\eex
\ensub
where $D(\cdot,\cdot)$ denotes some distance (eg., a trace norm) between the states,
$\delta > 0$ is a small parameter chosen a priori and 
$\xi^{\prime}_k \equiv {\rm Tr}_S[U\rho^{(k-1)}_S\otimes \xi U^{\dag}]$.

The interaction between a reservoir qubit and the system qubit is given by
the partial swap operation
$P(\eta) = (\cos\eta)\mathbb{I} + i(\sin\eta)S$, where $S$, the {\em swap} operator
acting on the state of two qubits is defined by 
$S|\psi\rangle\otimes|\phi\rangle = |\phi\rangle\otimes|\psi\rangle$.
It can be shown that $\eta$ can be chosen to enforce Eq. (\ref{secu}) according to
the relation:
\be
\sin\eta \le \sqrt{\delta/2}.
\ee
Thus the information contained in the unknown system state
is distributed in the correlations amongst the system and 
the reservoir qubits, whose marginal states are close to $\xi$.
As the authors  point out, this process can be used 
as a {\it quantum safe with a classical combination}.

Now we show how this particular feature can be turned into
a special case of the $(k_c,k_q,n,\mathbb{C})$ threshold scheme,
subject to the restriction that $\mathbb{Q} \subseteq \mathbb{C}$, so that
$k_q = q$, i.e. all q-players must be present to reconstruct the secret.
The homogenization is reversible and the original state 
of the system and the reservoir qubits can be unwound.
Perfect unwinding can be performed only when 
the system particle is correctly identified from among the 
$N+1$ output qubits, and it and the reservoir qubits
interact via the inverse of the original partial swap operation.
Therefore, in order to unwind the homogenized system, the classical 
information (denoted $K$) about the sequence of the qubit interactions 
is essential. Now, of the $(N+1)!$ possible orderings,
only one will reverse the original process.
The probability to choose the system qubit correctly is 
$1/(N+1)$. Even when the particle is choosen successfully,
there are still $N!$ different possibilities in choosing 
the sequence of interaction with the reservoir qubits.
Thus, without the knowledge of the correct ordering,
the probability of successfully unwinding the homogenization
transformation is $1/((N+1)!)$, which is exponentially small in $N$.
Moreover, a particular order of trial unwinding and measurement 
will irrecoverably destroy the system. This is demonstrated in Figures (4)
and (5) of Ref. \cite{zim02}, where various wrong permutations of the ordering,
chosen by trial-and-error strategy, are shown not to reproduce the state.
So, for sufficiently large value of $N$, no
information about the system qubit can be deduced without this classical
information. Nevertheless, it is worth noting that the above computational argument, while
rendering security of the homogenizing procedure intuitively understandable and
highly plausible, does not rigorously prove it, even in the $N \rightarrow \infty$ limit.

A secondary wall of security is provided by the smallness of $\delta$. 
It is useful if players already have knowledge of $\xi$. In this case,
because of conditions (\ref{secu}), the homogenized state of the system or
reservoir qubit cannot be distinguished from $\xi$.

If $K$ is split up among the $q$ members holding the system and
reservoir qubits according to a $(q,q)$-CTS,
it is easy to observe that this realizes a $(q,q)$-QTS 
not based directly on a quantum error-correction code. 
In terms of the generalized definition, this
corresponds to a $(k_c,k_q,n,\mathbb{C})$-scheme in which 
$k_c=0$, $\mathbb{Q} = \mathbb{C}$
and $n=k_q=q$.
The classical layer of information sharing is
necessary in order to strictly enforce the threshold: if prior ordering
information were openly available, then for example
the last $q-1$ participants could collude to obtain a state close to $\rho$.
We now present the most general twin-threshold scheme possible based
on homogenization. It will still be more restricted than that obtained via
quantum encryption, requiring that $\mathbb{Q} \subseteq \mathbb{C}$, so
that $k_q=q$. If $n$ is not
too large, it is preferable for prevention of partial information leakage
to choose the number $N$ of
reservoir qubits such that $N \gg n$. The general protocol is executed 
recursively as follows.

{\em Scheme 3}: Protocol to realize a restricted $(k_c,k_q,n,\mathbb{C})$-Q2TS+C,
with $k_q=q \le n$: Alice takes $N$ ($\gg 1$) reservoir qubits, 
where $N+1 = \sum_i m_i$ and integers $m_i \ge 1$ ($\forall~ i: 1 \le i\le n$), and 
performs the process of homogenization to obtain states $\xi_0,\xi_1,
\cdots \xi_N$ on the system qubit and the $N$ reservoir qubits.

{\em Distribution phase:} (1)
Any $m_i$ qubits from $N+1$ qubits are given to the $i$th member of 
$\mathbb{Q}$; (2) $K$ is divided into two parts, $K_1$ and $K_2$, according 
to a (2,2)-CTS;
(3) Let $\lambda_c \equiv |\bar{\mathbb{Q}} ~\cap~ \mathbb{C}| \ge 0$.
$K_1$ is further split among the members of $\mathbb{Q}$ 
and $\bar{\mathbb{Q}} ~\cap~ \mathbb{C}$ using a 
$(q+\lambda_c,q+\lambda_c)$-CTS; (4) $K_2$
is split among the members of $\bar{\mathbb{Q}}-\mathbb{C}$ using a 
$(k_c-\lambda_c,n-q-\lambda_c)$-CTS.

{\em Reconstruction phase:} (1) Collect all q-shares from members of $\mathbb{Q}$;
(2) Collect all $|\mathbb{C}|$ c-shares from members of $\mathbb{C}$ and 
reconstruct $K_1$; (3) Collect any $k_c-\lambda_c$ shares from members of
$\bar{\mathbb{Q}}-\mathbb{C}$ to reconstruct $K_2$; (4) Using $K_1$ and $K_2$,
reconstruct $K$; (5) Using the q-shares and $K$,
unwind the system state to restore the secret $|S\rangle$.

{\em Acknowledgments:}
We are grateful to Dr. Mueller-Quade for pointing out an important reference.
We thank Prof. Anil Kumar, Prof. J. Pasupathy and Mr. Ranabir Das for discussions.
SKS thanks Prof. Anil Kumar for enabling his visit to IISc during which this
work was done. RS's work, begun at the Center for Theoretical Studies,
was partially supported by DRDO project 510 01PS-00356.

\end{document}